\documentclass[12pt]{article}

\usepackage{comment}
\usepackage{apacite}
\usepackage{graphicx,psfrag,epsf}
\usepackage{enumerate}
\usepackage{natbib}
\usepackage{mathtools}
\mathtoolsset{showonlyrefs}
\usepackage{booktabs}
\usepackage{color}
\usepackage{subcaption}
\usepackage[english]{babel} % English language/hyphenation
\usepackage[protrusion=true,expansion=true]{microtype} % Better typography
\usepackage{amsmath,amsfonts,amsthm}
\usepackage{amssymb}
\usepackage{chngcntr}
\usepackage[toc,page]{appendix}
\usepackage{textcomp}
\usepackage{url}
\usepackage{bigints}
\usepackage{placeins}
\usepackage{array}
\usepackage{booktabs} % Horizontal rules in tables
\usepackage{setspace}
\usepackage{amsmath}
\usepackage{hyperref}
\usepackage{subcaption}
\usepackage{adjustbox}
\usepackage{yhmath}
\usepackage{pdfprivacy}
\usepackage{soul}
\usepackage{multirow}
\usepackage{enumitem}
\usepackage[font = small,labelfont=bf,textfont=it]{subcaption}
\usepackage[table]{xcolor}
\usepackage{diagbox}
\usepackage{tabularx}
\usepackage[font = small,labelfont=bf,textfont=it]{caption} % Custom captions under/above floats in tables or figures
\usepackage{footnote}
\usepackage{algorithm}% http://ctan.org/pkg/algorithms
\usepackage{algpseudocode}% http://ctan.org/pkg/algorithmicx

\usepackage{etoolbox}
\usepackage{tabularx}
\usepackage{authblk}
\usepackage{caption}
\usepackage{indentfirst}
\usepackage{enumitem}
\usepackage{makecell}
\newlist{steps}{enumerate}{1}
\setlist[steps, 1]{label = Step \arabic*:}
\usepackage{tikz}
\tikzstyle{process} = [rectangle, rounded corners, minimum width=2cm, minimum height=1cm, text width=2cm, text centered, draw=black, fill=white]
\tikzstyle{arrow} = [thick,->,>=stealth]
\usetikzlibrary{shapes,decorations,arrows,calc,arrows.meta,fit,positioning}
\tikzset{
    -Latex,auto,node distance =1 cm and 1 cm,semithick,
    state/.style ={ellipse, draw, minimum width = 0.7 cm},
    point/.style = {circle, draw, inner sep=0.04cm,fill,node contents={}},
    bidirected/.style={Latex-Latex,dashed},
    el/.style = {inner sep=2pt, align=left, sloped}
}
\tikzstyle{decision} = [diamond, 
minimum width=2cm, 
minimum height=1cm, 
text centered, 
draw=black,
fill=white]
\algblock{ParFor}{EndParFor}
\algnewcommand\algorithmicparfor{\textbf{for}}
\algnewcommand\algorithmicpardo{\textbf{do\ parallel}}
\algnewcommand\algorithmicendparfor{\textbf{end\ parallel\ for}}
\algrenewtext{ParFor}[1]{\algorithmicparfor\ #1\ \algorithmicpardo}
\algrenewtext{EndParFor}{\algorithmicendparfor}

\makeatletter
\def\BState{\State\hskip-\ALG@thistlm}

\newcommand{\distas}[1]{\mathbin{\overset{#1}{\kern\z@\sim}}}%

\newsavebox{\mybox}\newsavebox{\mysim}
\newcommand{\distras}[1]{%
  \savebox{\mybox}{\hbox{\kern3pt$\scriptstyle#1$\kern3pt}}%
  \savebox{\mysim}{\hbox{$\sim$}}%
  \mathbin{\overset{#1}{\kern\z@\resizebox{\wd\mybox}{\ht\mysim}{$\sim$}}}%
}
\newtheorem{theorem}{Theorem}

\newtheorem{proposition}{Proposition}

\newtheorem{definition}{Definition}

\newcommand{\be}{\begin{equation}}
\newcommand{\ee}{\end{equation}}
    \newcommand{\bi}{\begin{itemize}}
\newcommand{\ei}{\end{itemize}}
\newcommand{\ben}{\begin{enumerate}}
\newcommand{\een}{\end{enumerate}}

\newcommand*{\rom}[1]{\expandafter\@slowromancap\romannumeral #1@}

\newcolumntype{K}[1]{\geq {\centering\arraybackslash}p{#1}}
\allowdisplaybreaks

\makeatother

\let\oldbibliography\thebibliography
\renewcommand{\thebibliography}[1]{\oldbibliography{#1}
\setlength{\itemsep}{0pt}} %Reducing spacing in the bibliography.

\newcommand{\blind}{0}

\patchcmd{\footnotemark}{\stepcounter{footnote}}{\refstepcounter{footnote}}{}{}
\begin{document}

\def\spacingset#1{\renewcommand{\baselinestretch}%
{#1}\small\normalsize} \spacingset{1}

\if0\blind
{
  \title{\bf Efficient Screening Designs for Expensive Black-box Models with Qualitative and Quantitative Factors
  }
  \small
   \author{Difan Song\textsuperscript{1} and V. Roshan Joseph \textsuperscript{2} \vspace{3mm} \\
   \textsuperscript{1}Department of Statistics, Harvard University \vspace{1mm} \\
    \textsuperscript{2}H. Milton Stewart School of Industrial and Systems Engineering, \\
    Georgia Institute of Technology}
  \date{}
  \maketitle
} \fi

\if1\blind
{
 \title{\bf }
  \bigskip
  \bigskip
  \bigskip
  \begin{center}
    {\LARGE\bf Efficient Screening Designs for Expensive Black-box Models with Qualitative and Quantitative Factors}
\end{center}

  \medskip
  %\maketitle
} \fi

\begin{abstract}
Computationally expensive black-box models often involve a large number of input factors with complex interactions and varying importance. Experimental design techniques can be used for quickly identifying the important factors, which can make the optimization of a complex computer model or the training of an expensive machine learning model more efficient. Existing screening designs for black-box models focus mainly on continuous factors, with the maximum one-factor-at-a-time (MOFAT) design being a recent example. In this work, we extend the design to incorporate multiple types of factors, including nominal, ordinal, and discrete-numeric. We first identify the properties leading to optimal screening, where qualitative and quantitative factors should be treated differently. Based on these properties, we propose practical algorithms to efficiently construct MOFAT designs for all types of factors. The usefulness of the design is demonstrated by both numerical experiments and an application to hyperparameter tuning in machine learning models.
\end{abstract}

\vfill

\newpage
\spacingset{1.55} % DON'T change the spacing!

\section{Introduction} \label{sec:intro}

In today's digital world, computationally intensive black-box models help advance knowledge in various fields. Many of these models involve a large number of both qualitative and quantitative factors as inputs. For instance, in plasma physics, \citet{Knapp_Lewis_Joseph_Jennings_Glinsky_2023} describes a model for the output of plasma radiation detectors, where the outputs depend on filter material (qualitative) and filter thickness (quantitative). Another example is hyperparameter tuning of machine learning algorithms, such as decision trees \citep{pmlr-v133-turner21a} and deep learning \citep{MathWorksBayesianOpt}. In the latter case, the choice of activation function is qualitative, and many hyperparameters, such as the learning rate and regularization, are quantitative. The high computational cost of evaluating computer models or training machine learning models necessitates careful experimental design to explore the input space. The aim of this article is to develop experimental design techniques that can quickly search in a large input space involving both qualitative and quantitative factors and identify the few important factors that affect the output.

In the realm of physical experiments \citep{Wu2021}, most of the experimental designs, such as orthogonal arrays, are based on qualitative factors. Although there have been efforts in extending them to include quantitative factors \citep{cheng2004geometric, zhou2014space}, they consider only linear or quadratic effects, which are inadequate for screening in complex black-box models. The same deficiency is present in response surface designs and screening designs that focus on quantitative factors and their extensions to include qualitative factors \citep{wu1998construction, jones2013definitive}. On the other hand, in the literature on computer experiments \citep{joseph2026experimental}, there are many designs that incorporate both qualitative and quantitative factors. Sliced Latin hypercube designs (SLHDs; \citealp{Qian_2012, Ba_Myers_Brenneman_2015}), marginally coupled designs (MCDs; \citealp{Deng_Hung_Lin_2015}), and doubly coupled designs (DCDs; \citealp{Yang_Lin_Zhou_He_2023}) all constrain the quantitative factors to have a Latin hypercube structure for levels or combinations of levels of the qualitative factors. MaxPro design \citep{Joseph_Gul_Ba_2020} offers an alternative with more flexible run sizes and better projection properties. 

However, we note that none of the above designs allows for model-free identification of important factors. In complicated models, we want to understand the nonlinear effects and complex interactions of the factors, but without relying on any specific statistical or machine learning modeling technique. When only continuous factors are involved, Morris screening \citep{Morris_1991} achieves this goal by using a collection of one-factor-at-a-time (OFAT) designs. These designs are widely used in the field of sensitivity analysis \citep{sensitivitybook2021}. \citet{Xiao_Joseph_Ray_2023} recently proposed maximum one-factor-at-a-time (MOFAT) designs as an improvement to the Sobol' design \citep{sobol1993sensitivity}, recognizing the latter as a special class of OFAT designs.

In this work, we extend the idea of MOFAT designs to incorporate all types of factors. Most of the factors found in real applications can be classified into four types: nominal (qualitative), ordinal (qualitative), discrete-numeric (quantitative), and continuous (quantitative). Since ordinal factors can be converted into discrete-numeric factors using the scoring method \citep[Ch.12]{Wu2021}, we focus on the desired properties and construction methods for nominal and discrete-numeric factors for the remainder of this work.

A main motivation for extending MOFAT designs comes from the recent finding of \citet{Song_Joseph} that they are excellent choices for initial designs in active learning. Random Latin hypercube designs or other space-filling designs are commonly used as initial designs in active learning methods such as Bayesian optimization \citep{pourmohamad2021bayesian}. \citet{Song_Joseph} found that MOFAT designs outperform the state-of-the-art initial designs because of their superior ability to estimate length-scale parameters in Gaussian process models. Therefore, it is necessary to enhance the flexibility of MOFAT designs to accommodate different types of factors to fully realize their potential in real applications.

The article is organized as follows. Section \ref{sec:background} reviews screening designs for continuous factors and summarizes the three properties that guide the construction of MOFAT designs. Sections \ref{sec:nominal} and \ref{sec:discrete_numeric} extend the properties to nominal and discrete-numeric factors, respectively, and provide construction algorithms. Section \ref{sec:space_filling} discusses projection issues for MOFAT designs, and an algorithm to overcome them. Section \ref{sec:simulation} compares the screening performance of the proposed design with that of the existing designs through simulations. Section \ref{sec:application} applies the design to a machine learning hyperparameter tuning problem. We conclude with some remarks in Section \ref{sec:conclusion}.

\section{Background: MOFAT designs for continuous factors} \label{sec:background}

This section provides background on screening designs for deterministic black-box models with continuous factors, which we assume to lie in $[0, 1]^p$ (after re-scaling) throughout this text. In contrast with physical experiments, where the main and quadratic effects are of primary interest, nonlinear effects and complex interactions are often of interest in expensive, deterministic black-box models.

\subsection{Morris and Sobol' Designs}
One popular choice for screening designs is the Morris screening design \citep{Morris_1991}, which is based on a collection of $l$ OFAT designs randomly placed on the grid $\{0, 1/(l - 1), \dots, 1\}^p$. \citet{Xiao_Joseph_Ray_2023} recently made a connection between Morris screening and pick-freeze sampling methods for estimating the total Sobol' indices \citep{sobol1993sensitivity, homma1996importance}. For a given number of base points $l$ and dimension $p$, a Sobol' design is constructed by: \begin{enumerate}
    \item Generate two $l \times p$ random designs on $[0, 1]^p$, $\boldsymbol{A}$ and $\boldsymbol{B}$.
    \item For $k = 1, \dots, p$, replace the $k$-th column of $\boldsymbol{A}$ with the $k$-th column of $\boldsymbol{B}$ to generate $\boldsymbol{A}^{(k)}$.
    \item Stack $\boldsymbol{A}, \boldsymbol{A}^{(1)}, \dots, \boldsymbol{A}^{(p)}$ together to get an $l(p + 1) \times p$ matrix as the design $\mathcal{D}$.
\end{enumerate} Thus, compared with the base points in $\boldsymbol{A}$, each point in $\boldsymbol{A}^{(k)}, k = 1, \dots, p$ only varies by one factor. Denoting the input factors as $\mathbf{x}$ and the output as $y$, the OFAT structure makes it straightforward to estimate the total Sobol' indices 
\begin{align}
t_k = \frac{\mathbb{E}_{\mathbf{x}_{\sim k}}\{\mathrm{Var}_{x_{k}}[y | \mathbf{x}_{\sim k}]\}}{\mathrm{Var}[y]}, k = 1, \dots, p, 
    \label{eqn:T_Sobol}
\end{align} using Monte Carlo methods \citep{Jansen_1999}: \begin{align}
    \hat t_k = 
    \frac{\frac{1}{2l} \sum_{i = 1}^l \left[y(\mathbf{A}_{i}) - y(\mathbf{A}_{i}^{(k)})\right]^2}{\widehat{\mathrm{Var}}(y)} = \frac{\widehat{V}_k^{\text{tot}}}{\widehat{\mathrm{Var}}(y)}, \label{eqn:T_Sobol_est}
\end{align} where $\sim k$ denotes all factors other than $k$, $\mathbf{A}_i$ and $\mathbf{A}_{i}^{(k)}$ denote the $i$-th row of $\boldsymbol{A}$ and $\boldsymbol{A}^{(k)}$, respectively. When there is no ambiguity, we also use $\mathbf{A}_k$ and $\mathbf{A}_{k}^{(k)}$ to denote the $k$-th column of $\boldsymbol{A}$ and $\boldsymbol{A}^{(k)}$, respectively. The total Sobol' index $t_k$ measures the contribution of factor $k$ to the variance of the output, which includes the main effect and all the interaction effects involving $x_k$. For expensive black-box models, this sensitivity measure is more suitable than derivative-based or model-based measures; hence, it will be the main focus of this work.
% As \cite{Xiao_Joseph_Ray_2023} points out, Morris screening can be viewed as a Sobol' design with a specific choice of $\boldsymbol{A}$ and $\boldsymbol{B}$.

\subsection{MOFAT Designs}
\citet{Xiao_Joseph_Ray_2023} proposed an improvement of the Sobol' design and named their designs as maximum one-factor-at-a-time (MOFAT) designs. The matrix $\boldsymbol{A}$ containing the base points is chosen as a maximin Latin hypercube design \citep{Morris_Mitchell_1995} that maximizes the $\ell_1$ distance, while the elements of $\boldsymbol{B}$ are given by $B_{ik} = T_l(A_{ik})$, where \begin{align}
    T_l(i) = i - \lfloor l/2 \rfloor + l \mathbb{I}\{i < (1 + l)/2\}, i = 1, \dots, l, \label{eqn:MOFAT_trans}
\end{align} where $\lfloor i \rfloor$ is the largest integer not exceeding $i$ and $\mathbb{I}\{\cdot\}$ denotes the indicator function. Here, the elements of $\boldsymbol{A}$ and $\boldsymbol{B}$ are in $\{1,2,\ldots,l\}$, which can be scaled to obtain the final design in $[0,1]^p$.

We summarize three properties in the construction of MOFAT designs with continuous factors that were not explicitly stated in the original work of \citet{Xiao_Joseph_Ray_2023}: \begin{enumerate}
    \item (OFAT) $A_{ik} \neq A_{ik}^{(k)}, A_{ik} = A_{ik}^{(\sim k)}$.

    \item (Uniformity) $\mathbf{A}_{k}$ and $\mathbf{A}_{k}^{(k)}$ are both uniform on $[0, 1]$.

    \item (MOFAT) $\sum_{i = 1}^l |A_{ik} - A_{ik}^{(k)}|$ is maximized for all $k = 1, \dots, p$.
\end{enumerate} The first property states that only one factor changes at a time, which is the definition of OFAT designs. While straightforward, the property reflects the famous aphorism ``no causation without manipulation'' \citep{Rubin_1978}. By the OFAT structure, we can attribute the variation in the outcome to specific factors. The second property enforces one-dimensional uniformity, which ensures the representativeness of the points, allowing for better estimation of the outer integral in \eqref{eqn:T_Sobol} with sample average in \eqref{eqn:T_Sobol_est} \citep[Ch.6]{joseph2026experimental}. Finally, the MOFAT criterion maximizes the sensitivity of the Sobol's index estimates, as shown by the following proposition from \citet{Xiao_Joseph_Ray_2023}:
\begin{proposition}
If the response surface $y(\mathbf{x})$ is a realization of a Brownian motion field \citep{Zhang_Apley_2014} with \begin{align}
    \mathbb{E}\left\{y(\mathbf{x}_1) - y(\mathbf{x}_2)\right\} = 0, \mathrm{Var}\left\{y(\mathbf{x}_1) - y(\mathbf{x}_2)\right\} = \sigma^2 \|\mathbf{x}_1 - \mathbf{x}_2\|_{\omega},
\end{align} where $\|\cdot\|_{\omega}$ is a weighted Euclidean distance on $[0, 1]^p$, then maximizing $\sum_{i = 1}^l |A_{ik} - A_{ik}^{(k)}|$ also maximizes the expected value of $\widehat{V}_k^{\text{tot}}$ for $k = 1, \dots, p$.
\end{proposition}
In other words, MOFAT designs maximize the numerator of Equation \eqref{eqn:T_Sobol_est} in expectation, making it a suitable choice for efficient factor screening. By the development of \citet{Xiao_Joseph_Ray_2023}, the following result holds:
\begin{proposition}
For continuous factors on $[0, 1]^p$, the construction given by the transformation \eqref{eqn:MOFAT_trans} satisfies all three properties.
\end{proposition}

In this work, we extend the above properties to nominal and discrete-numeric factors and develop algorithms to efficiently construct designs for these types of factors. We also provide an additional improvement to the projection of all types of factors. Our implementation of MOFAT designs with qualitative and quantitative factors is available in the open-source R package \texttt{MOFAT}.

\section{MOFAT designs for nominal factors} \label{sec:nominal}

\subsection{Maximization property for nominal factors}

In this section, we first discuss the properties for building a MOFAT design for nominal factors. For a nominal factor indexed by $k$, assume it has levels represented by $1, \dots, m_k$. The first two properties are straightforward. The OFAT property requires that the base point (a row of $\boldsymbol{A}$) and the derived point (corresponding row of $\boldsymbol{A}^{(k)}$) have different levels in factor $k$, while the uniformity property states that each level in $\{1, \dots, m_k\}$ has an equal number of occurrences, which is achievable when we let the number of base points $l$ be a multiple of $m_k$.

However, the third property requires additional attention. One choice is to use the Hamming distance for the distance between different levels: \begin{align}
    d_{\text{Hamming}}(\mathbf{a}, \mathbf{b}) = \sum_{i = 1}^l \mathbb{I}\{a_i \neq b_i\},
\end{align} where $\mathbb{I}\{\cdot\}$ denotes the indicator function. Then, any design satisfying the OFAT condition trivially maximizes the distance between $\mathbf{A}_{k}$ and $\mathbf{A}_{k}^{(k)}$. However, we often believe different nominal levels have varying degrees of similarity to each other. One popular modeling approach to capture the similarity is the latent variable Gaussian process (LVGP, \citealp{Zhang_Tao_Chen_Apley_2020}). While this is not the only approach for modeling nominal factors, we show that adopting this view yields an interesting yet intuitive property for the design. 

The latent variable model assumes that each level corresponds to a vector in a continuous latent space through a mapping $\mathbf{z}^{k}: \{1, \dots, m_k\} \to \mathbb{R}^2$. Following \citet{Zhang_Apley_2014}, we can assume that the response surface is a Brownian motion on the latent space: \begin{align}
    \mathbb{E}\left\{y[\mathbf{z}^k(x_1)] - y[\mathbf{z}^k(x_2)]\right\} = 0, \mathrm{Var}\left\{y[\mathbf{z}^k(x_1)] - y[\mathbf{z}^k(x_2)]\right\} = \sigma^2 \|\mathbf{z}^k(x_1) - \mathbf{z}^k(x_2)\|_{\gamma},
\end{align} where $\|\cdot\|_{\gamma}$ is now a weighted Euclidean distance on the latent space $\mathbb{R}^2$. For the numerator of the estimated total Sobol' indices \eqref{eqn:T_Sobol_est}: \begin{align*}
\mathbb{E}[\widehat{V}_k^{\text{tot}}] &= \frac{1}{2l} \sum_{i = 1}^l \mathbb{E}\left[y(\mathbf{A}_i) - y(\mathbf{A}_i^{(k)})\right]^2 \\
&= \frac{1}{2l} \sum_{i = 1}^l \mathrm{Var}\left[y(\mathbf{A}_i) - y(\mathbf{A}_i^{(k)})\right] \\
&= \frac{1}{2l} \sum_{i = 1}^l \mathrm{Var}\left\{y[\mathbf{z}^k(A_{ik})] - y[\mathbf{z}^k(A_{ik}^{(k)})]\right\} \\
&= \frac{\sigma^2}{2l} \sum_{i = 1}^l \left\|\mathbf{z}^k(A_{ik}) - \mathbf{z}^{k}(A_{ik}^{(k)})\right\|_{\gamma}.
\end{align*} The MOFAT design intends to maximize this expectation to increase the screening sensitivity. Unfortunately, we have no knowledge of $\mathbf{z}^k(\cdot)$ prior to the experiment. We instead define the following robust optimization problem: \begin{gather} \label{eqn:robust_opt}
\begin{array}{rl}
\max_{\mathbf{A}_{k}, \mathbf{A}_{k}^{(k)}} \min_{\left(\mathbf{z}^k(1), \dots, \mathbf{z}^k(m_k)\right)' \in \mathcal{Z}^k} & \sum_{i = 1}^l \left\|\mathbf{z}^k(A_{ik}) - \mathbf{z}^{k}(A_{ik}^{(k)})\right\|_{\gamma}, \\
\text{subject to} & A_{ik}, A_{ik}^{(k)} \in \{1, \dots, m_k\} \\
& A_{ik} \neq A_{ik}^{(k)}, i = 1, \dots, l.
\end{array}
\end{gather} The inner minimization problem is over an ambiguity set $\mathcal{Z}^k$, for which we assume to obey: \begin{align}
    \mathcal{Z}^k = \left\{(\mathbf{z}^k(1), \dots, \mathbf{z}^k(m_k))' \bigg| c \leq \sum_{1 \leq s < t \leq m_k}\left\|\mathbf{z}^k(s) - \mathbf{z}^k(t)\right\|_{\gamma} \leq C\right\},
\end{align} where $c < C$ are suitable constants. The ambiguity set constrains the sum of all pairwise distances in the latent space to reflect the belief that the levels are at least distinguishable to some extent, while not being too far to be completely uncorrelated. This assumption is an example of a permutation-invariant set reflecting the vagueness of our knowledge before the experiment, and the idea of which can be dated back to the work of \citet{Wu_1981}. 

To further simplify the robust optimization problem \eqref{eqn:robust_opt}, we collapse the sum of distances according to the levels of $A_{ik}$ and $A_{ik}^{(k)}$. For $1 \leq s < t \leq m_k$, define \begin{align}
    w_k(s, t) = \mathrm{card}\left\{i \in \{1, \dots, l\} \Big| A_{ik} = s, A_{ik}^{(k)} = t \text{ or } A_{ik} = t, A_{ik}^{(k)} = s\right\}, \label{eqn:cardinality}
\end{align} which counts the number of OFAT pairs with level combination $(s, t)$. Let $\mathbf{w}_k = \{w_{k}(s, t) | 1 \leq s < t \leq m_k\}$ denote the collection of counts, we can rewrite \eqref{eqn:robust_opt} as: \begin{gather} \label{eqn:robust_opt_2}
\begin{array}{rl}
\max_{\mathbf{w}_k} \min_{\left(\mathbf{z}^k(1), \dots, \mathbf{z}^k(m_k)\right)' \in \mathcal{Z}^k} & \sum_{1 \leq s < t \leq m_k} w_k(s, t) \left\|\mathbf{z}^k(s) - \mathbf{z}^{k}(t)\right\|_{\gamma}, \\
\text{subject to} & \sum_{1 \leq s < t \leq m_k} w_k(s, t) = l.
\end{array}
\end{gather}

To characterize the optimal solution of \eqref{eqn:robust_opt_2}, we first introduce the definition of majorization, and interested readers can refer to the monograph by \citet{marshall2011inequalities}.
\begin{definition}
For $\mathbf{a}, \mathbf{b} \in \mathbb{R}^n$, $\mathbf{a}$ is said to be majorized by $\mathbf{b}$ (or $\mathbf{b}$ majorizes $\mathbf{a}$), denoted as $\mathbf{a} \prec \mathbf{b}$ (or $\mathbf{b} \succ \mathbf{a}$), if \begin{align}
\begin{cases}
\sum_{i=1}^{k} a_{[i]} \leq \sum_{i=1}^{k} b_{[i]}, & k = 1, \ldots, n-1, \\
\sum_{i=1}^{n} a_{[i]} = \sum_{i=1}^{n} b_{[i]},
\end{cases}
\end{align} where the subscript $[i]$ denotes the $i$-th largest element.
\end{definition}
Majorization provides a partial order for sorted vectors. Intuitively, $\mathbf{a}$ is majorized by $\mathbf{b}$ if the elements of $\mathbf{a}$ are distributed more uniformly. With this notion, we are ready to state the optimality result: 
\begin{theorem} \label{theorem:majorization}
The optimal solution of \eqref{eqn:robust_opt_2} is the vector $\mathbf{w}_k$ that is majorized by all other vectors. The result also holds if we replace the Brownian motion assumption by a Gaussian process with zero mean and a correlation function that is nonnegative and decreasing.
\end{theorem}

The proof of the theorem uses the fact that the objective function is \textit{Schur-concave}. In words, we prefer designs where the level combinations in OFAT pairs appear an equal number of times. This result shows that without prior knowledge of the separation in the latent space, we should explore each possibility equally for robustness, which matches our intuition. Hence, we state the result as the MOFAT property for nominal factors: \begin{enumerate}
    \item[3\textquotesingle.] (MOFAT for nominal factors) For all $k$ corresponding to nominal factors, the distribution of the counts $w_k(s, t)$ is majorized by all other possible distributions subject to $\sum_{1 \leq s < t \leq m_k} w_k(s, t) = l$.
\end{enumerate}

We emphasize that although we used the latent variable approach \citep{Zhang_Tao_Chen_Apley_2020} as a motivation, the resulting property is general and exists independent of the modeling approach. During the data analysis stage, one can use any preferred approach on the collected data. 

\subsection{Construction of MOFAT designs for nominal factors}

With the properties (1), (2), and (3') in mind, we state our algorithm for constructing MOFAT designs for nominal factors. The first step is constructing the base design $\mathbf{A}_k$ of size $l$. Starting from the values of any Latin hypercube design with values $\{1, \dots, l\}$, we apply the following pointwise transformation: \begin{align}
    U_{kl}(i) = \lfloor (i - 1) m_k/ l \rfloor + 1, \quad i = 1, \dots, l, \label{eqn:U_{kl}}
\end{align} where $\lfloor \cdot \rfloor$ denotes the floor function. This transformation ensures that the output is in $\{1, \dots, m_k\}$, and each level appears the same number of times when $l \bmod m_k = 0$, where $\bmod$ denotes the modular operation which gives the remainder after division.

Next, we need another transformation to map the levels to different levels in $\mathbf{B}_k$. The pointwise transformation that we propose is: \begin{align} \label{eqn:V_{kl}}
    V_{kl}(i) &= \{[(i - 1) \bmod (l/m_k) \bmod (m_k - 1)] + U_{kl}(i)\} \bmod m_k + 1, \quad i = 1, \dots, l.
\end{align}

For a better understanding of this seemingly complicated transformation, we present some examples in Table \ref{tab:nominal}. For a given $l$ and $m_k$, $U_{kl}(\cdot)$ simply divides the base points into blocks of size $l/m_k$, and assigns the levels $1, \dots, m_k$ to each block. $V_{kl}(\cdot)$ then transforms $U_{kl}(x)$ to values in $\{1, \dots, m_k\} \backslash U_{kl}(x)$. For example, when $l = 9, m_k = 3$, $U_{kl}(\cdot)$ transforms the first three numbers to level 1. Then, the corresponding outputs of $V_{kl}(\cdot)$ are the levels 2 and 3, repeating in cycles ($2, 3, 2, 3, \dots$).

\begin{table}[tb]
\centering
\caption{Examples of levels of nominal factors in the base design $\mathbf{A}_k$ given by $U_{kl}(\cdot)$ and $\mathbf{B}_k$ given by $V_{kl}(\cdot)$ for different values of $l$ and $m_k$.}
\label{tab:nominal}
\begin{tabular}{@{}lccccc@{}}
\toprule
& $l = 6, m_k = 2$ & & $l = 6, m_k = 3$ & & $l = 9, m_k = 3$ \\ \midrule
LHD & $1\ 2\ 3\ 4\ 5\ 6$ & & $1\ 2\ 3\ 4\ 5\ 6$ & & $1\ 2\ 3\ 4\ 5\ 6\ 7\ 8\ 9$ \\
$\mathbf{A}_k$ & $1\ 1\ 1\ 2\ 2\ 2$ & & $1\ 1\ 2\ 2\ 3\ 3$ & & $1\ 1\ 1\ 2\ 2\ 2\ 3\ 3\ 3$ \\
$\mathbf{B}_k$ & $2\ 2\ 2\ 1\ 1\ 1$ & & $2\ 3\ 3\ 1\ 1\ 2$ & & $2\ 3\ 2\ 3\ 1\ 3\ 1\ 2\ 1$ \\ \bottomrule
\end{tabular}
\end{table}

The following result formally states that the design constructed using $U_{kl}(\cdot)$ and $V_{kl}(\cdot)$ satisfies the previously discussed properties. A proof is provided in the supplementary material.
\begin{proposition} \label{proposition:nominal}
Assume we have a nominal factor $k$ with levels $1, \dots, m_k$. Let $b_1, \dots, b_l$ be a permutation of $\{1, \dots, l\}$, with $l \bmod m_k = 0$. For $i = 1, \dots, l$, let \begin{align}
\begin{cases}
    A_{ik} = U_{kl}(b_i), \\
    A_{ik}^{(k)} = V_{kl}(b_i),
\end{cases}
\end{align} with $U_{kl}(\cdot)$ and $V_{kl}(\cdot)$ given in \eqref{eqn:U_{kl}} and \eqref{eqn:V_{kl}}, respectively. Then the design satisfies: 
\begin{enumerate}
    \item (OFAT) $A_{ik} \neq A_{ik}^{(k)}, A_{ik} = A_{ik}^{(\sim k)}$.

    \item (Uniformity) $\mathbf{A}_{k}$ and $\mathbf{A}_{k}^{(k)}$ are both uniform on $\{1, \dots, m_k\}$.

    \item[3\textquotesingle.] (MOFAT) The distribution of $w_k(s, t)$ defined in \eqref{eqn:cardinality} is majorized by all other possible distributions.
\end{enumerate}
\end{proposition}

\section{MOFAT for discrete-numeric factors} \label{sec:discrete_numeric}

Suppose for a discrete-numeric factor $k$ with $m_k$ levels, the (distinct) factor levels in increasing order are $a_1 < \dots < a_{m_k}$. Due to the discrete and quantitative nature of the factor, we can take the uniformity property from nominal factors and the MOFAT property from continuous factors. However, we face a new difficulty when maximizing $\sum_{i = 1}^l |A_{ik} - A_{ik}^{(k)}|$. Unlike continuous and nominal factors, where we can place the values on a regular grid, the values $\{a_1, \dots, a_{m_k}\}$ can be arbitrary. Hence, we cannot always easily construct a design that achieves maximization through pre-defined transformations such as \eqref{eqn:MOFAT_trans}. 

Instead, we propose to formulate finding the OFAT pairs as a weighted bipartite matching problem. Formally, given two sets $\mathcal{L} = \mathcal{R} = \{a_1, \dots, a_{m_k}\}$ and a cost function $\xi: \mathcal{L} \times \mathcal{R} \to \bar{\mathbb{R}}$, we want to find a bijection $f$ such that $\sum_{i = 1}^{m_k} \xi(a_i, f(a_i))$ is minimized. In our problem, we can define the cost function to be \begin{align} \label{eqn:cost_matrix}
\xi(a_i, a_j) = \begin{cases}
    (a_{m_k} - a_1) - |a_i - a_j| & a_i \neq a_j, \\
    \infty & a_i = a_j.
\end{cases} 
\end{align} This cost function ensures that (1) all costs are nonnegative; (2) the cost decreases linearly in the distance between two levels; and (3) pairs with $a_i = a_j$ are never chosen. We can also represent the costs in matrix form, and Figure \ref{fig:Hungarian} presents an example with $(a_1, a_2, a_3, a_4) = (1, 2, 3, 5)$. 

\begin{figure}[tb]
\centering
\begin{tikzpicture}[
% Global options for the picture
node distance = 0.5cm and 0.5cm, % vertical and horizontal distance between nodes
]

\node[align=center] (A) {
Iteration 1 \\
$\begin{array}{c | cccc}
\text{levels}  & 1 & 2 & 3 & 5 \\ \hline
1 & \cdot & 3 & \cellcolor{cyan} 2 & 0 \\
2 & 3 & \cdot & 3 & \cellcolor{cyan} 1 \\
3 & 2 & \cellcolor{cyan} 3 & \cdot & 2 \\
5 & \cellcolor{cyan} 0 & 1 & 2 & \cdot
\end{array}$
};

\node[right=of A] (next1) {$\longrightarrow$};

\node[right=of next1, align=center] (B) {
Iteration 2 \\
$\begin{array}{c | cccc}
\text{levels}  & 1 & 2 & 3 & 5 \\ \hline
1 & \cdot & 3 & \cdot & \cellcolor{cyan} 0 \\
2 & 3 & \cdot & \cellcolor{cyan} 3 & \cdot \\
3 & \cellcolor{cyan} 2 & \cdot & \cdot & 2 \\
5 & \cdot & \cellcolor{cyan} 1 & 2 & \cdot
\end{array}$
};

\node[right=of B] (next2) {$\longrightarrow$};

\node[right=of next2, align=center] (C) {
Iteration 3 \\
$\begin{array}{c | cccc}
\text{levels}  & 1 & 2 & 3 & 5 \\ \hline
1 & \cdot & \cellcolor{cyan} 3 & \cdot & \cdot \\
2 & \cellcolor{cyan} 3 & \cdot & \cdot & \cdot \\
3 & \cdot & \cdot & \cdot & \cellcolor{cyan} 2 \\
5 & \cdot & \cdot & \cellcolor{cyan} 2 & \cdot
\end{array}$
};

\node[below=of B] (D) {
% $\begin{array}{cccc}
% 1 & 2 & 3 & 5 \\ \hline
% 3 & 5 & 2 & 1 \\
% 5 & 3 & 1 & 2 \\
% 2 & 1 & 5 & 3
% \end{array}$

$\begin{array}{cccc} \toprule
& \multicolumn{3}{c}{l = 12, (a_1, a_2, a_3, a_4) = (1, 2, 3, 5)} \\ \midrule
& \text{Iteration 1} & \text{Iteration 2} & \text{Iteration 3} \\
\mathbf{A}_k & 1 \ 2 \ 3 \ 5 & 1 \ 2 \ 3 \ 5 & 1 \ 2 \ 3 \ 5 \\
\mathbf{B}_k & 3 \ 5 \ 2 \ 1 & 5 \ 3 \ 1 \ 2 & 2 \ 1 \ 5 \ 3 \\ \bottomrule
\end{array}$
};

\node[left=of D] (result) {$\Longrightarrow$};

\end{tikzpicture}
\caption{An illustration of the Hungarian algorithm. Top: Cost matrices and the solutions found by three iterations of the Hungarian algorithm. A dot indicates the entry is not allowed to be chosen (has cost $\infty$). Bottom: The base design $\mathbf{A}_k$ and $\mathbf{B}_k$ corresponding to the solutions.}
\label{fig:Hungarian}
\end{figure}

\begin{algorithm}[tb]
\caption{Applying the Hungarian algorithm to find OFAT designs} \label{alg:Hungarian}
\begin{algorithmic}[1]
\Require Base design $\mathbf{A}_k \in \{a_1, \dots, a_{m_k}\}^{l}$, procedure \texttt{HungarianSolver}, number of procedure iterations $\bar h$
\State Initialize cost matrix $\mathcal{C} \in \mathbb{R}^{m_k \times m_k}$, array $\texttt{count} = [\underbrace{0, \dots, 0}_{m_k}]$;
\State $\mathcal{C}_{ij} \gets (a_{m_k} - a_1) - |a_{i} - a_j|$ for $1 \leq i, j \leq m_k$;
\State $\mathcal{C}_{ii} \gets \sum_{i \neq j}\mathcal{C}_{ij} + 1$ for $i = 1, \dots, m_k$;

\For{$h \gets 1 \text{ to } \bar h$}
\State $f_h \gets \texttt{HungarianSolver}(\mathcal{C})$;
\State $\mathcal{C}_{i, j \text{ such that } f_h(a_i) = a_j} \gets \sum_{i \neq j} \mathcal{C}_{ij} + 1$ for $i = 1, \dots, m_k$;
\EndFor

\For{$i \gets 1 \text{ to } l$}
\State $B_{ik} \gets f_{(\texttt{count}[A_{ik}] \bmod \bar h) + 1}(A_{ik})$;
\State \texttt{count}[$A_{ik}$] += 1;
\EndFor
\State \Return Vector {$\mathbf{B}_k$};
\end{algorithmic}
\end{algorithm}

Having defined the cost matrix, it is straightforward to solve the optimization problem using the Hungarian algorithm \citep{kuhn1955hungarian}. The top-left panel in Figure \ref{fig:Hungarian} shows the result of applying the algorithm for the first time. The solution yields the bijection $f_1$ that maps $(1, 2, 3, 5)$ to $(3, 5, 2, 1)$. This will constitute four OFAT designs. In this example with $l = 12$ base points, each level appears three times in the base design, so we need three bijections in total. Furthermore, it is desirable that each $a_i$ is mapped to different levels, increasing the diversity of the design. Therefore, we can assign a cost of $\infty$ to the selected entries, and reapply the Hungarian algorithm to find the next bijection $f_2$ that gives four more OFAT designs. The process is similar for acquiring $f_3$ and the last set of OFAT designs.

We summarize the procedure in Algorithm \ref{alg:Hungarian}, where we make use of the subprocedure \texttt{HungarianSolver} that returns the desired bijection given the cost matrix. An implementation can be found in the R package \texttt{RcppHungarian} \citep{RcppHungarian}. We keep track of how many times each level has appeared in the base design, and loop through the solutions $f_1, \dots, f_h$ for the desired OFAT design. Finally, we formally state that the algorithm returns a design that satisfies our three properties.
\begin{proposition} \label{proposition:discrete_num}
Assume that we have a discrete-numeric factor $k$ with levels $a_1 < \dots < a_{m_k}$. Let $b_1, \dots, b_l$ be a permutation of $\{1, \dots, l\}$, with $l \bmod m_k = 0$. For $i = 1, \dots, l$, let \begin{align}
\begin{cases}
    A_{ik} = a_{U_{kl}(b_i)}, \\
    A_{ik}^{(k)} = B_{ik} \text{ returned by Algorithm \ref{alg:Hungarian}},
\end{cases}
\end{align} with $U_{kl}(\cdot)$ given in \eqref{eqn:U_{kl}} and $\bar h = \lfloor \frac{m_k^2 + 3}{2m_k}\rfloor$ in Algorithm \ref{alg:Hungarian}. Then the design satisfies: 
\begin{enumerate}
    \item (OFAT) $A_{ik} \neq A_{ik}^{(k)}, A_{ik} = A_{ik}^{(\sim k)}$.

    \item (Uniformity) $\mathbf{A}_{k}$ and $\mathbf{A}_{k}^{(k)}$ are both uniform on $\{a_1, \dots, a_{m_k}\}$.

    \item (MOFAT) $\sum_{i = 1}^l |A_{ik} - A_{ik}^{(k)}|$ is maximized.
\end{enumerate}
\end{proposition}

We make a remark on the requirement that $l \bmod m_k = 0$ for both nominal and discrete-numeric factors. Although we need this condition to satisfy the uniformity property, it is not a strict requirement for the \texttt{mofatQQ} function in the \texttt{MOFAT} package. Our implementation accepts any $l \geq 3$, which further enhances run-size flexibility. We recommend $l \geq \max_{k} m_k$ in practice so that each level appears at least once in the base design.

\section{Projections of MOFAT designs} \label{sec:space_filling}

All three properties stated in the previous sections describe the behavior of a single factor. In this section, we discuss the space-fillingness of MOFAT designs with qualitative and quantitative factors. In particular, we consider the projections in 2 to $p$ dimensions, which is important for understanding the nonlinear and interaction effects in the subspaces of the influential factors.

\begin{figure}[tb]
\centering
\includegraphics[width=0.8\textwidth]{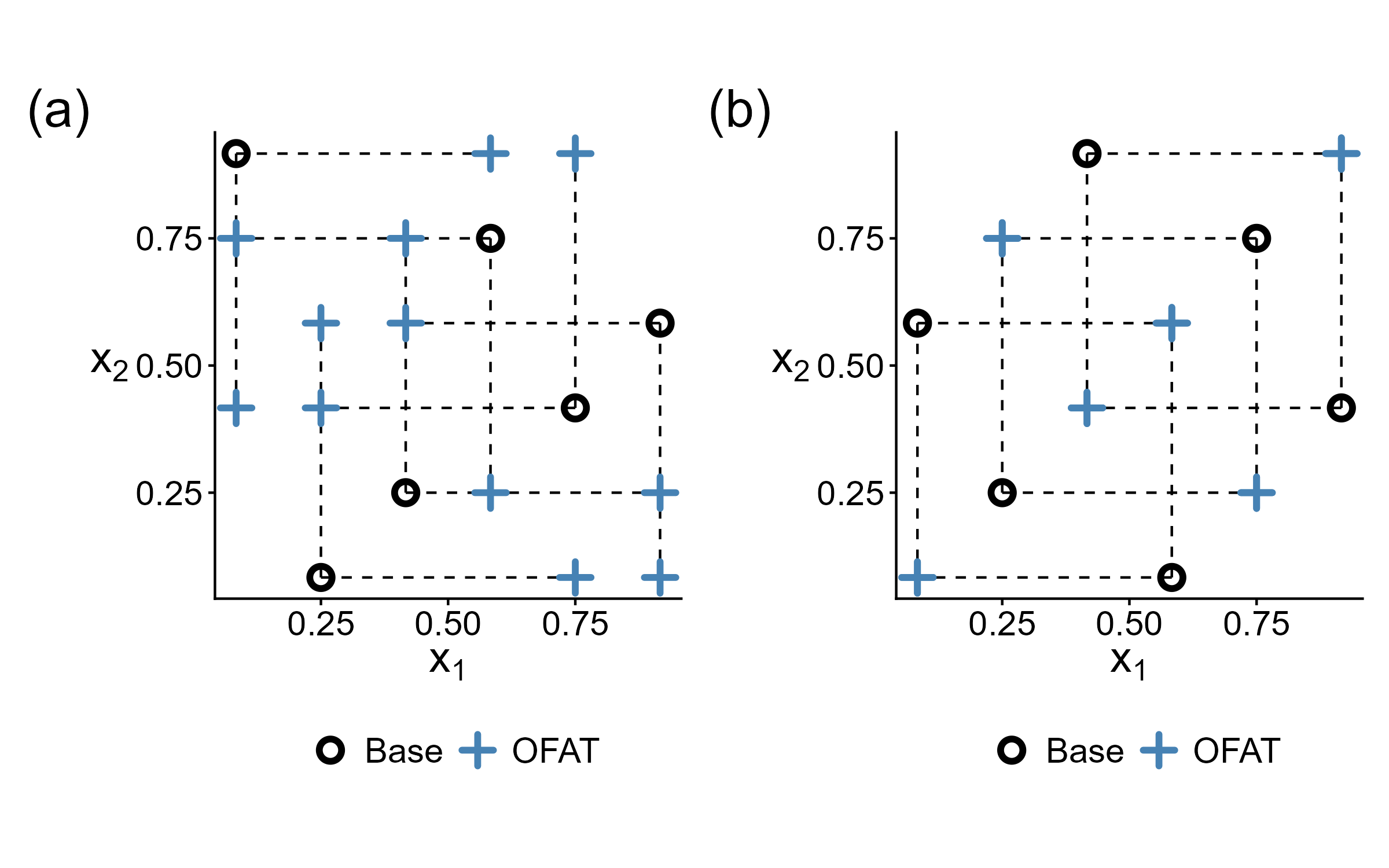}
\caption{Visualization of 2d projections (continuous factors) for a 10d MOFAT design. (a) Case with good 2d projection. (b) Case with bad 2d projection. Dashed lines show pairs of OFAT designs.}
\label{fig:projection}
\end{figure}

Proposition 3 of \citet{Xiao_Joseph_Ray_2023} states that MOFAT designs with at least three columns do not have duplicated rows. Nevertheless, there will be duplicates in the subspaces due to the OFAT structure. Take the example in Figure \ref{fig:projection}, which shows the projection in the first two dimensions for two designs with $l = 6$ and $p = 10$. Because $\boldsymbol{A}^{(3)}, \dots, \boldsymbol{A}^{(10)}$ are the same as $\boldsymbol{A}$ in the first two dimensions, we can have at most $3l = 18$ unique points in the 2d subspace. The left panel of Figure \ref{fig:projection} shows such an example, which fills the $[0, 1]^2$ region well. The right panel shows another example with a different base design, but now there are only 12 unique points, leaving larger gaps in the design region and making the design unfavorable. 

Without changing the construction algorithms described previously, we propose to improve the projections by a multistart scheme: we generate multiple designs and select the optimal design according to some criterion. We use Gower's dissimilarity \citep{Gower_1971} to incorporate different types of factors: \begin{align}
    \bar d(x_{ik}, x_{jk}) = \begin{cases}
        \mathbb{I}\{x_{ik} \neq x_{jk}\} & \text{factor $k$ is nominal}, \\
        \frac{|x_{ik} - x_{jk}|}{\max \mathbf{x}_{k} - \min \mathbf{x}_{k}} & \text{otherwise}.
    \end{cases}
\end{align} Let $\mathcal{K} \subseteq \{1, \dots, p\}$ denote a subset of dimension indices. Motivated by the example in Figure \ref{fig:projection}, we want to avoid having duplicates in the 2d subspaces, or rather, maximize the minimum separation for all $\mathcal{K}$ with size two: \begin{align}
    \lambda^* = \min_{\mathrm{card}\{\mathcal{K}\} = 2} \min_{\substack{1 \leq i < j \leq n \\ G(i), G(j) \in \{0\}\cup \mathcal{K}}} \sum_{k \in \mathcal{K}} \bar d(x_{ik}, x_{jk}), \label{eqn:minimum_separation}
\end{align} where, $n = l(p + 1)$ is the design size, and the index \begin{align}
    G(i) = \begin{cases}
        0 & \mathbf{x}_i \in \boldsymbol{A}, \\
        k & \mathbf{x}_i \in \boldsymbol{A}^{(k)}, k = 1, \dots, p
    \end{cases}
\end{align} indicates the block membership of a row $\mathbf{x}_i$, and the condition $G(i), G(j) \in \{0\}\cup \mathcal{K}$ removes the duplicates of the base design. Although we care about projections in subspaces of more than two dimensions, we provide a result showing that considering \eqref{eqn:minimum_separation} is sufficient.
\begin{proposition} \label{proposition:projection}
Let the value of \eqref{eqn:minimum_separation} be $\lambda^*$. For the base design, define \begin{align}
    \lambda_{\boldsymbol{A}}(q) = \min_{\mathrm{card}\{\mathcal{K}\} = q} \min_{1 \leq i < j \leq l}\sum_{k \in \mathcal{K}} \bar d(x_{ik}, x_{jk}), q = 1, \dots, p.
\end{align} Then for the MOFAT design, we have that \begin{align}
    \min_{\mathrm{card}\{\mathcal{K}\} = q} \min_{\substack{1 \leq i < j \leq n \\ G(i), G(j) \in \{0\}\cup \mathcal{K}}} \sum_{k \in \mathcal{K}} \bar d(x_{ik}, x_{jk}) \geq  \lambda_{\boldsymbol{A}}(q - 2) + \lambda^*
\end{align} for all $3 \leq q \leq p$.
\end{proposition} 
The proposition shows that once the base design is fixed, the separations of the MOFAT design in all subspaces are bounded by the minimum separation in 2d. This property is a result of the standard OFAT structure, which ensures that when comparing two different points, at most two coordinates are not from the base design.

For practical implementation, we can let the base design $\boldsymbol{A}$ be a MaxPro design \citep{Joseph_Gul_Ba_2015} that ensures good projection properties in all subspaces. Then, for the full MOFAT design, instead of maximizing \eqref{eqn:minimum_separation}, we optimize the reciprocal criterion following \citet{Morris_Mitchell_1995}: \begin{align}
    \min_{\mathcal{D}} \left\{\sum_{\mathrm{card}\{\mathcal{K}\} = 2} \sum_{\substack{1 \leq i < j \leq n \\ G(i), G(j) \in \{0\}\cup \mathcal{K}}} \frac{1}{\sum_{k \in \mathcal{K}} {\bar d}^2(x_{ik}, x_{jk}) + 1/l^2}\right\}, \label{eqn:final_criterion}
\end{align} where the term $1/l^2$ is added in the denominator to prevent the value from becoming infinite. The expression also has a close connection with the criterion in \citet{Draguljić_Santner_Dean_2012}, which incorporates projections in all subspaces: \begin{align}
    \min_{\mathcal{D}} \left\{\sum_{q = 1}^p \sum_{\mathrm{card}\{\mathcal{K}\} = q} \sum_{1 \leq i < j \leq n} \frac{q}{\sum_{k \in \mathcal{K}} {\bar d}^2(x_{ik}, x_{jk})}\right\}.
\end{align} Due to the difficulty of computing all combinations, \citet{Draguljić_Santner_Dean_2012} focused on subspaces with $q \leq 2$. In the specific case of the MOFAT design, we have shown through Proposition \ref{proposition:projection} that projections with $q \geq 3$ are controlled. Furthermore, the one-dimensional projections are guaranteed by the uniformity property. Hence, only $q = 2$ enters our final criterion \eqref{eqn:final_criterion}.

\begin{figure}[tb]
\centering
\includegraphics[width=0.8\textwidth]{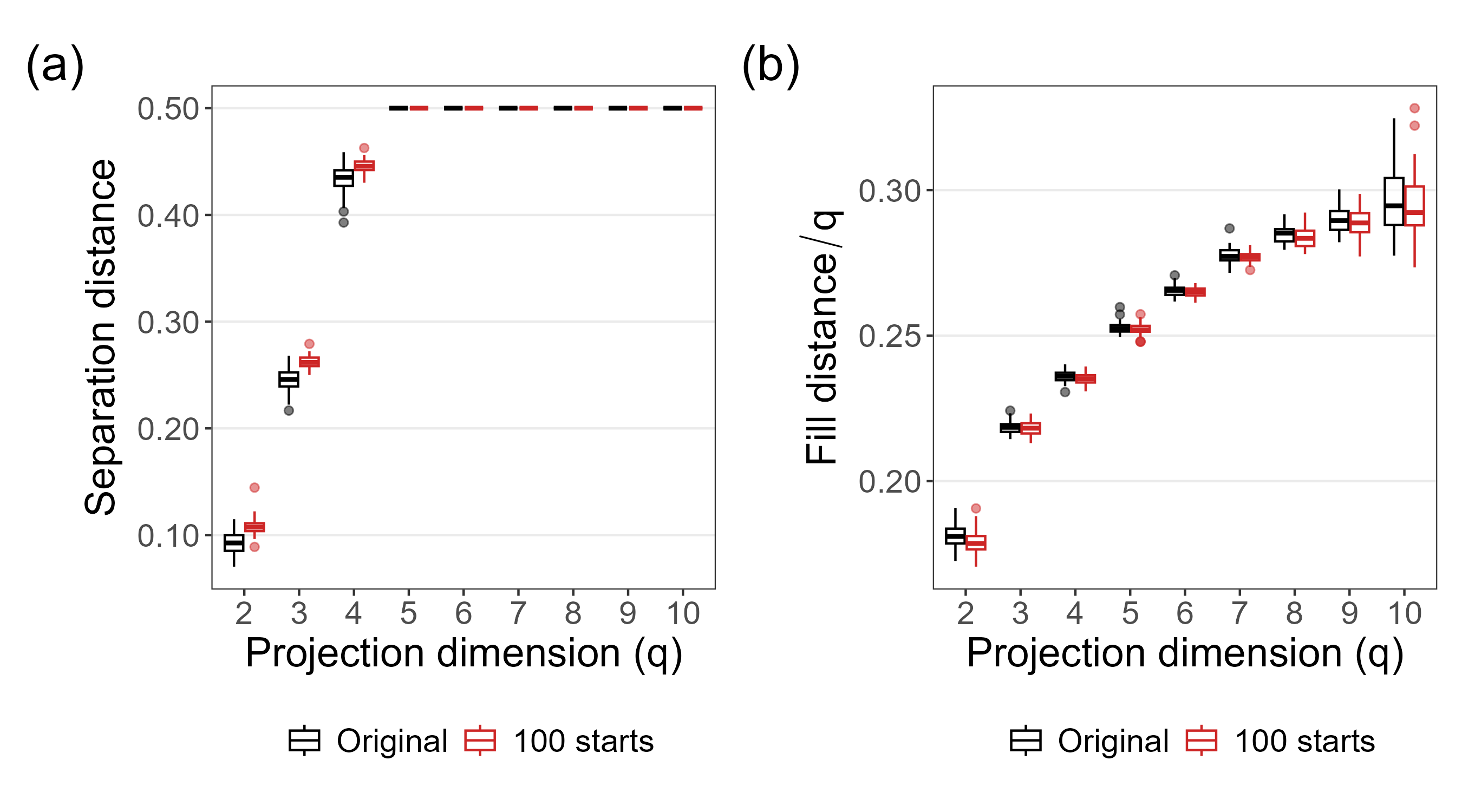}
\caption{Relationship of the criterion with two space-filling criteria in projected spaces. (a) Average separation distance. (b) Average fill distance. The designs are for ten continuous factors, so the fill distance can be evaluated.}
\label{fig:criteria}
\end{figure}

Finally, we show a visualization comparing our criterion with two criteria averaged on all $\mathcal{K}$ with $\mathrm{card}\{\mathcal{K}\} = q$: \begin{align}
    \text{Separation distance: } & \frac{1}{\binom{p}{q}} \sum_{\mathrm{card}\{\mathcal{K}\} = q} \min_{\substack{1 \leq i < j \leq n \\ G(i), G(j) \in \{0\}\cup \mathcal{K}}} \sum_{k \in \mathcal{K}} \bar d(x_{ik}, x_{jk}), \\
    \text{Fill distance: } & \frac{1}{\binom{p}{q}} \sum_{\mathrm{card}\{\mathcal{K}\} = q} \max_{\mathbf{u} \in [0, 1]^q} \min_{i:\, G(i) \in \{0\}\cup \mathcal{K}} \|\mathbf{u} - \mathbf{x}_{i, \mathcal{K}}\|_1.
    % \text{Energy distance: } & \left\{2\bar{\mathbb{E}}[\|\mathbf{x}_{i, \mathcal{K}} - \mathbf{u}\|_1] -  \bar{\mathbb{E}}[\|\mathbf{x}_{i, \mathcal{K}} - \mathbf{x}_{j, \mathcal{K}}\|_1] - \bar{\mathbb{E}}[\|\mathbf{u}_i - \mathbf{u}_j\|_1]\right\}^{1/2}, \\
    % & \hspace{-5em} \text{where $\bar{\mathbb{E}}[\cdot]$ denotes the empirical mean, and $\mathbf{u}$ is a uniform sample on $[0, 1]^2$.}
\end{align} Figure \ref{fig:criteria} shows an example with ten continuous factors, so that we can calculate the fill distance. Nevertheless, the results are qualitatively similar when other types of factors are included. First, the separation distance (larger the better) is the motivation for using criterion \eqref{eqn:final_criterion}. From panel (a), multiple starts optimizing the criterion indeed improve the separation distances in low dimensions. Although for $q \geq 5$, the separation no longer improves, panel (b) shows that the fill distance (smaller the better) improves in all subspaces. The cost of improving projections by multistart is also modest in our implementation (the ten-dimensional example runs in about one second for 100 starts).

\section{Numerical experiment} \label{sec:simulation}

In this section, we demonstrate the use of the proposed MOFAT designs for the emulation of the Borehole function, which models water flow through a borehole: \begin{align}
f(\mathbf{x})=\frac{2 \pi T_u\left(H_u-H_l\right)}{\ln \left(r / r_w\right)\left(1+\frac{2 L T_u}{\ln \left(r / r_w\right) r_w^2 K_w}+\frac{T_u}{T_l}\right)}. \label{eqn:Borehole}
\end{align} See \citet{simulationlib} for specific ranges of the factors. Following \citet{Zhou_Qian_Zhou_2011}, we take three factors $(r, H_u - H_l, K_w)$ and discretize each of them to three equally spaced values. During the simulation, we treat these three factors as nominal, while it is also possible to treat them as discrete-numeric. Therefore, we have four continuous factors and three nominal factors from the model. For screening purposes, we also add six continuous factors and seven nominal factors with three levels that are inert, resulting in $p = 20$ factors in total. Because all nominal factors have three levels, the smallest MOFAT design for this example has $l = 3$ and $n = l(p + 1) = 63$.

Part of our motivation for the MOFAT design was that no existing designs were specialized for screening qualitative and quantitative factors. Nevertheless, we can adapt the Sobol' design described in Section \ref{sec:background} to create a design for comparison. Namely, we randomly sample $\boldsymbol{A}$ and $\boldsymbol{B}$ on the design space, with the constraint that $A_{ik} \neq B_{ik}$ for all $i, k$. Stacking $\boldsymbol{A}, \boldsymbol{A}^{(1)}, \dots, \boldsymbol{A}^{(p)}$, where $\boldsymbol{A}^{(k)}$ replaces the $k$-th column of $\boldsymbol{A}$ with the $k$-th column of $\boldsymbol{B}$, we have a Sobol' design that shares the same structure as the MOFAT design. It satisfies the OFAT property, which is beneficial for screening, but lacks uniformity, MOFAT, or improved space-filling properties.

In addition, we compare with MCD \citep{Deng_Hung_Lin_2015} and MaxPro designs \citep{Joseph_Gul_Ba_2020} of the same size. As we described in Section \ref{sec:intro}, these designs incorporate qualitative and quantitative factors, but are not targeted at screening. We do not compare with SLHDs or DCDs because an SLHD requires a Latin hypercube slice for all level combinations of the nominal factors; a DCD requires the number of nominal factors to be no more than the number of factor levels. These restrictions make the designs prohibitive when the number of nominal factors is large. For this example with ten nominal factors with three levels, by Proposition 4 of \citet{Deng_Hung_Lin_2015}, MCDs exist and can be constructed from mixed orthogonal arrays $OA(63, 3^{10} 21, 2)$. On the other hand, MaxPro designs exist for any $n$.

Given a design $\mathcal{D} = \{\mathbf{x}_i\}_{i = 1}^n$ and noise-free outputs $\mathbf{y}$, we model the data with a Gaussian process (GP) prior: \begin{align}
    f(\mathbf{x}) \sim \mathcal{GP}(\mu, \sigma^2 R(\mathbf{x}, \mathbf{x}')),
\end{align} where $R(\cdot, \cdot)$ is a positive-semidefinite correlation kernel function. We are particularly interested in nonnegative anisotropic kernels of the form \begin{align}
    R(\mathbf{x}, \mathbf{x}') = K\left(\sqrt{\sum_{k} \frac{(x_k - x_k')^2}{\theta_k^2}}\right),
\end{align} with examples including the squared exponential, rational quadratic, and Mat{\'e}rn class kernels \citep[Ch.4]{williams2006gaussian}. Under the GP assumption, the observed data have a multivariate Gaussian likelihood (up to additive constants): \begin{align}
    \log L(\mu, \sigma^2, \boldsymbol{\theta}) = -\frac{1}{2} \left[n \log \sigma^2 + \log |\boldsymbol{R}| + \frac{1}{\sigma^2} (\mathbf{y} - \mu \mathbf{1})' \boldsymbol{R}^{-1}(\mathbf{y} - \mu \mathbf{1}) \right], \label{eqn:GP_likelihood}
\end{align} where $\boldsymbol{R} = [R(\mathbf{x}_i, \mathbf{x}_j)]_{i, j = 1}^n \in [0, 1]^{n \times n}$ is the correlation matrix at the observations, and $\mathbf{1}$ is a vector of ones. We can obtain maximum likelihood estimates of $(\mu, \sigma^2, \boldsymbol{\theta})$ by numerically maximizing \eqref{eqn:GP_likelihood}.

For the experiments in Sections \ref{sec:simulation} and \ref{sec:application}, we perform Helmert encoding \citep{Helmert} for nominal factors, which works as follows: for a factor with $m_k$ levels, construct the $m_k \times (m_k - 1)$ normalized Helmert matrix with entries \begin{align}
    H_{ij} = \begin{cases}
        -1/\sqrt{j(j + 1)}, & i \leq j, \\
        \sqrt{j/(j + 1)}, & i = j + 1, \qquad j = 1, \dots, m_k - 1, \\
        0, & i \geq j + 2
    \end{cases}
\end{align} and level $i$ is encoded as row $i$ of $\boldsymbol{H}$. It is straightforward to verify that $\boldsymbol{H}' \mathbf{1} = \mathbf{0}$, $\boldsymbol{H}' \boldsymbol{H} = \boldsymbol{I}$, and the Euclidean distance between any two distinct row vectors is $\sqrt{2}$, ensuring the levels are equidistant. Also, the numeric factors are scaled to $[0, 1]$. This encoding allows us to use standard correlation kernels for all factors. We use the squared-exponential kernel, while also noting that our results are robust to this choice.

\begin{figure}[tb]
\centering
\includegraphics[width=\textwidth]{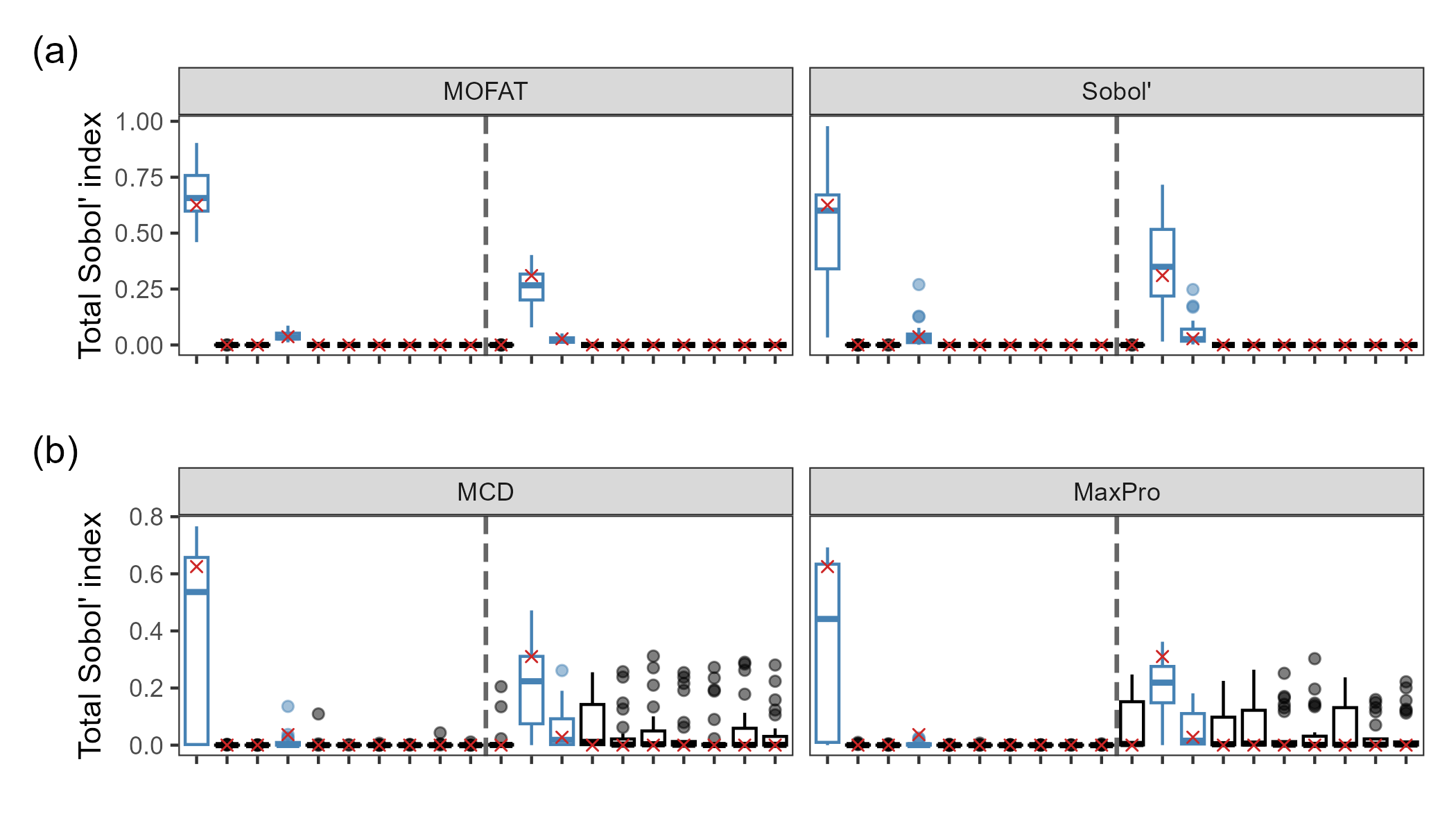}
\caption{Estimated Sobol' indices for the Borehole function with quantitative and qualitative inputs. (a) (Normalized) total Sobol' indices for screening designs. (b) (Normalized) total Sobol' indices for space-filling designs. The crosses show the Sobol'-Jansen estimator on the original function. All designs have 63 points and are repeatedly generated 25 times with different seeds. In each plot, the colored bars show the influential factors. On the left are quantitative factors, on the right are qualitative factors.}
\label{fig:Borehole}
\end{figure}

We first compare the estimated total Sobol' indices from each design. As described in Section \ref{sec:background}, MOFAT and Sobol' designs allow for Monte Carlo estimates using Equation \eqref{eqn:T_Sobol_est} directly. For MCD and MaxPro, we cannot directly use the estimator due to the lack of OFAT structure in the designs. However, it is possible to obtain sensitivity indices by utilizing the fitted GP surrogate. We can evaluate a large number of OFAT pairs on the surrogate and build a Sobol'-Jansen type estimator of the total Sobol' indices.

Panel (a) of Figure \ref{fig:Borehole} shows the estimated total Sobol' indices using \eqref{eqn:T_Sobol_est} for MOFAT and the Sobol' design, while panel (b) shows the Sobol'-Jansen estimator on the surrogate for MCD and MaxPro. We also estimate the ``true'' values by a Sobol'-Jansen estimator with 10,500 evaluations on the original function, and show them as crosses in the plots. For better visual comparison, we normalize the estimators to sum to one. The colored bars are the influential factors $(r_w, L, H_u - H_l, K_w)$ which have variance contributions greater than $1\%$. The identified factors also agree with the result of \citet{moon2010design}. From the plots, we can see that MOFAT and Sobol' designs both identify the important factors. The main difference is that MOFAT estimates are generally less volatile and closer to the true value, showing the benefits of building a more principled design. In contrast, MCD and MaxPro designs have much larger uncertainty for the important factors. They also produce non-negligible estimates for the inert factors, especially the nominal factors that have fewer unique levels than the continuous factors, which is undesirable.

\begin{figure}[tb]
\centering
\includegraphics[width=0.4\textwidth]{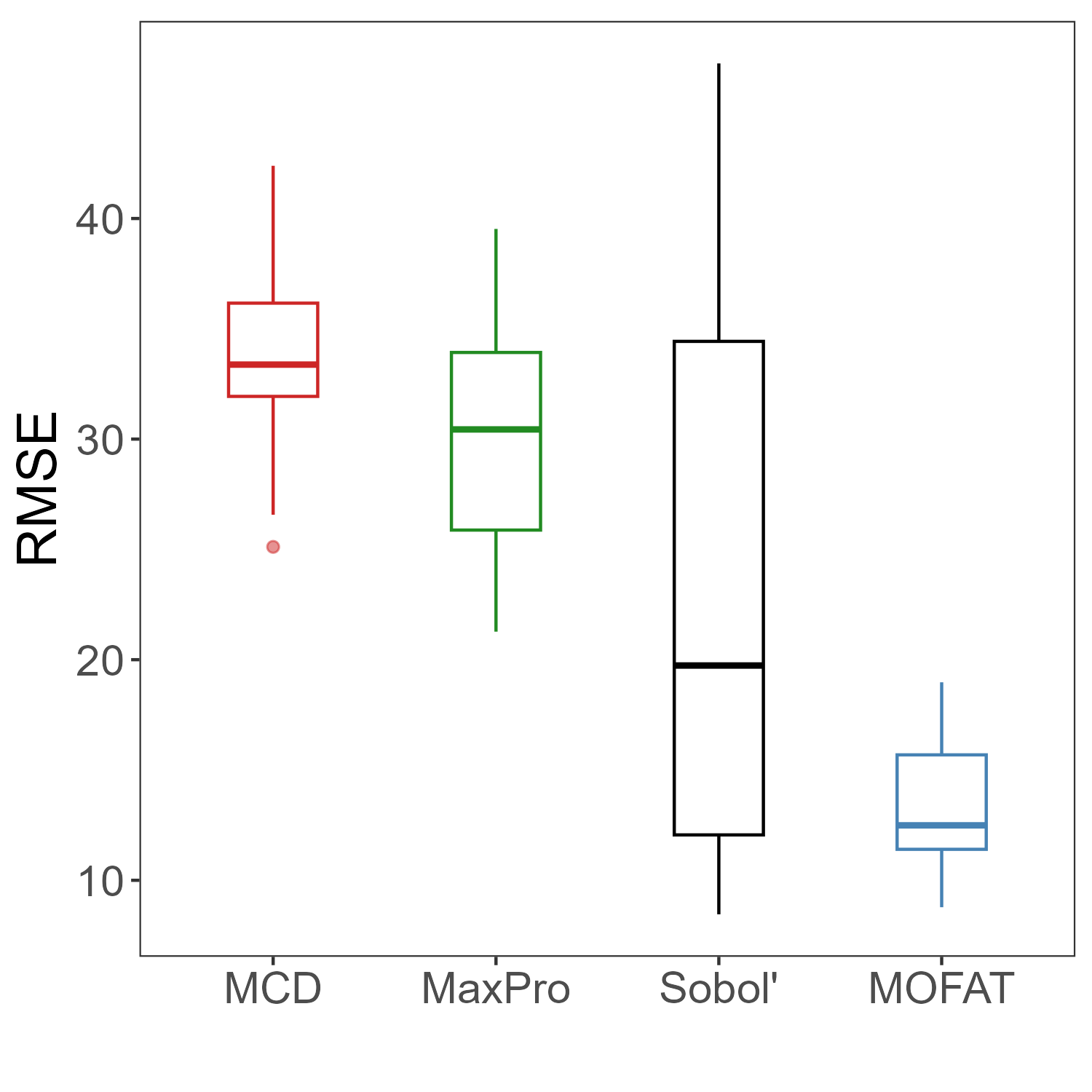}
\caption{Root mean square error (RMSE) of surrogates fitted on different designs. For each design, the surrogate is fit on factors with non-zero estimated Sobol' indices. All designs have 63 points and are repeatedly generated 25 times with different seeds.}
\label{fig:Borehole_RMSE}
\end{figure}

A second comparison we make is the quality of the fitted surrogate. We have discussed in Section \ref{sec:space_filling} that MOFAT designs trade off space-filling properties for screening capabilities. However, screening also helps improve the quality of the GP surrogate, because it is easier to fit a model in a lower dimension. For this reason, we take a screen-then-fit approach: take only the factors with non-zero estimated Sobol' indices from the previous step, and then fit the GP surrogate on those factors. Figure \ref{fig:Borehole_RMSE} presents the out-of-sample root mean square error (RMSE) on a fresh sample of 1000 test points. We found that MOFAT and Sobol' designs always select six factors (all factors except $T_u$ in the original Borehole function), while MCD and MaxPro select 14 to 18 factors in our simulations. Therefore, screening designs not only identify the important factors, but also benefit the screen-then-fit procedure in turn.

In the supplementary material, we provide the results for another physical model, the wing weight function, for which the results also demonstrate that MOFAT improves screening without compromising modeling performance. We conclude that MOFAT, as a specialized screening design with improved space-filling and optimality properties, is the most efficient choice for factor screening.

\section{Application: hyperparameter tuning} \label{sec:application}

\begin{table}[tb]
\centering
\caption{LightGBM hyperparameters and their search spaces.}
\label{tab:hyperparameters}
\begin{tabular}{llll}
\toprule
Name & Description & Type & Range/Values \\
\midrule
learning\_rate     & Learning rate (shrinkage)                & continuous         & [0.01, 0.2] \\
bagging\_fraction  & Row subsampling fraction                 & continuous         & [0.3, 1.0] \\
lambda\_l1         & $\ell_1$ regularization                        & continuous         & [0, 10] \\
lambda\_l2         & $\ell_2$ regularization                        & continuous         & [0, 10] \\
num\_leaves*        & Maximum leaves per tree                & integer            & [8, 32] \\
min\_data\_in\_leaf* & Minimum data in a leaf          & integer            & [2, 20] \\
num\_iterations    & Number of trees & discrete-numeric      & 100, 200, 300, 400, 500, 1000 \\
max\_depth         & Maximum tree depth                       & discrete-numeric     & 3, 6, 9, 12, 15, 20 \\
bagging\_freq      & Bagging frequency & discrete-numeric     & 1, 2, 3, 5, 10, 20 \\
boosting           & Boosting algorithm                       & nominal       & gbdt, rf, dart \\
tree\_learner      & Tree learner & nominal        & serial, voting \\
\bottomrule
\multicolumn{4}{p{\textwidth}}{\footnotesize{*\texttt{num\_leaves} and \texttt{min\_data\_in\_leaf} must be integers, so technically, they should be treated as discrete-numeric. However, since they can take many values, we treat them as continuous during the design stage and round them before inputting into \texttt{LightGBM}.}}
\end{tabular}
\end{table}

In this section, we present an application of the MOFAT design with qualitative and quantitative factors for hyperparameter tuning in machine learning models. Our demonstration uses \texttt{LightGBM} \citep{ke2017lightgbm}, which is an efficient implementation of gradient boosting decision trees. The official documentation of \texttt{LightGBM} describes a suite of tuning parameters that affect the quality of the fitted model\footnote{\url{https://lightgbm.readthedocs.io/en/latest/Parameters.html\#}}. Here, we take 11 hyperparameters, of which two are nominal, three are treated as discrete-numeric, and six are treated as continuous. See Table \ref{tab:hyperparameters} for details on the hyperparameters and the search space.

We use the Superconductivity data available at the UCI Machine Learning Repository \citep{superconductivty_data_464}. The dataset contains 21,263 superconductors, and the goal is to use the 81 extracted features to predict the critical temperature (the temperature below which a material loses all electrical resistance). Gradient boosted decision trees are well-suited for this high-dimensional prediction problem. However, as the sample size and input dimension become large, the computational cost of training and cross-validation for a set of hyperparameters also increases. Hence, there is a need to identify the ideal hyperparameter settings using limited cross-validation runs.

\begin{figure}[tb]
\centering
\includegraphics[width=0.5\linewidth]{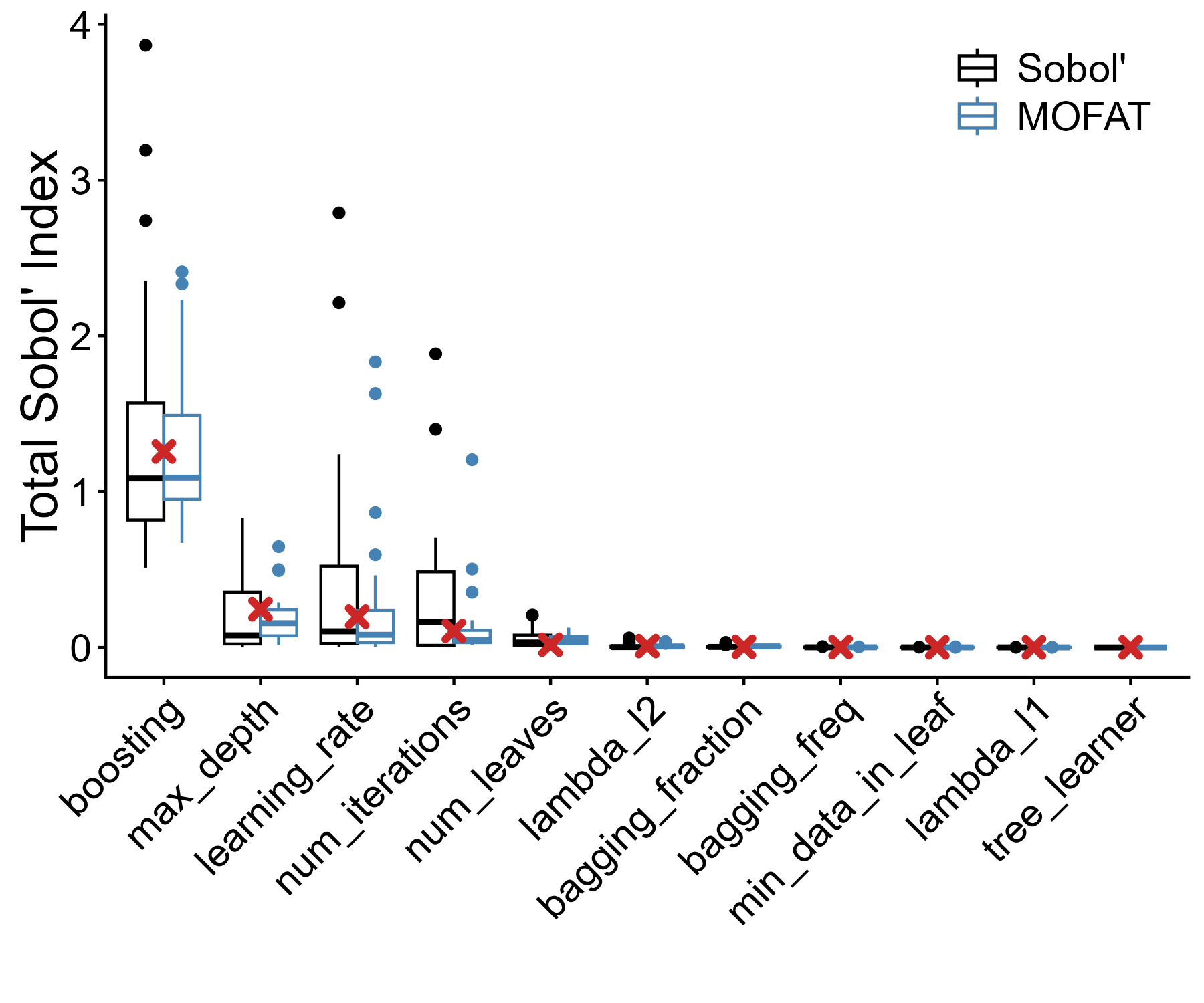}
\caption{Sobol' indices of hyperparameters. The boxplots are generated from 25 different designs of size 72. The red cross is the Sobol'-Jansen estimator from 600 evaluations.}
\label{fig:tree_sobol}
\end{figure}

We randomly select 5,000 rows as the training set and use the five-fold cross-validation mean-squared error on the training set as our objective function. As long as the folds are fixed, the objective function is noiseless in the sense that the same hyperparameters always give the same cross-validation error. For MOFAT, taking $l = 6$ gives 72-point designs on the hyperparameters. Similar to Section \ref{sec:simulation}, we generate Sobol' and MOFAT designs over 25 different seeds, and plot the estimated total Sobol' indices in Figure \ref{fig:tree_sobol}. We also evaluate the total Sobol' indices \eqref{eqn:T_Sobol_est} on a larger set of 600 different hyperparameter settings as reference values. The top four most influential hyperparameters are the boosting algorithm (nominal), maximum tree depth (discrete-numeric), learning rate (continuous), and number of trees (discrete-numeric), showing the significance of incorporating all factor types into the design. Furthermore, the hyperparameters show a clear hierarchy in importance, highlighting the value of factor screening. In this case, the MOFAT design again produces estimates that are more concentrated around the reference values, reliably informing the importance rankings of the tunable hyperparameters for this supervised learning task.

\begin{figure}[tb]
\centering
\includegraphics[width=0.5\textwidth]{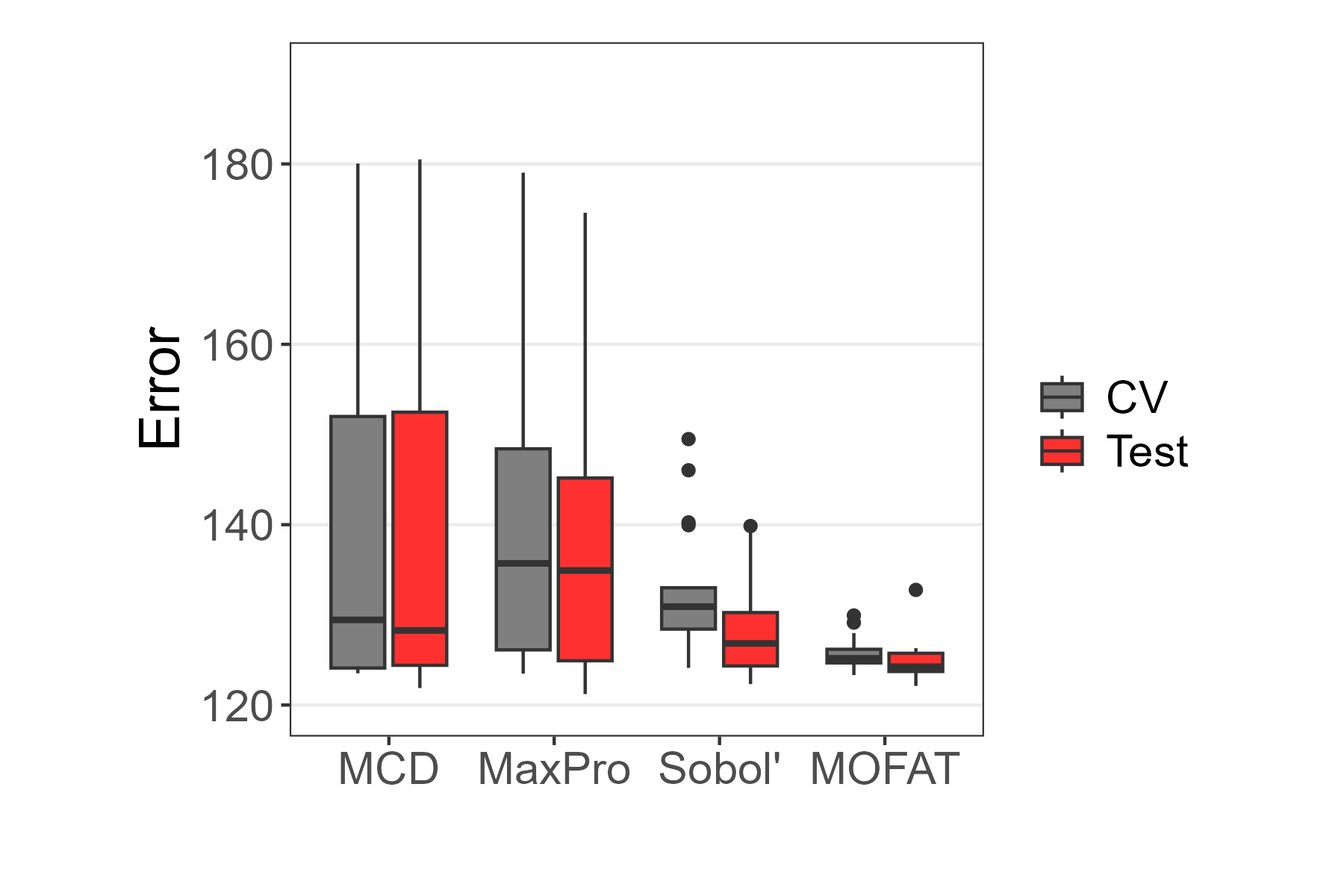}
\caption{Cross-validation errors and test errors for all hyperparameter settings proposed during sequential design.}
\label{fig:CV_test_error}
\end{figure}

To find hyperparameter settings that minimize cross-validation error, it is common to apply \textit{active learning}, which sequentially proposes new points based on the available information and uncertainty quantification. For this purpose, we again fit a GP surrogate. Then, we select the new hyperparameter setting as the maximizer of the expected improvement (EI) criterion \citep{Jones_Schonlau_Welch_1998}, which balances exploration and exploitation: \begin{align}
    \mathrm{EI}(\mathbf{x}) = \mathbb{E}[\max\{y_n^* - y(\mathbf{x}), 0\}], \label{eqn:EI_def}
\end{align} where $y_n^*$ is the minimum cross-validation error so far. Under the GP prior assumption, the posterior of $y(\mathbf{x})$ is also Gaussian. Denote this distribution by $\mathcal{N}(\hat{y}(\mathbf{x}), s^2(\mathbf{x}))$, and the expected improvement has a closed-form expression: \begin{align} \label{eqn:EI}
    \mathrm{EI}(\mathbf{x}) = \{y_n^* - \hat y(\mathbf{x})\} \Phi(u(\mathbf{x})) + s(\mathbf{x}) \phi(u(\mathbf{x})), 
    \text{ where } u(\mathbf{x}) = \frac{y_n^* - \hat y(\mathbf{x})}{s(\mathbf{x})},
\end{align} where $\Phi(\cdot)$ and $\phi(\cdot)$ denote the distribution and density functions of the standard Gaussian, respectively. We continue this process and sequentially select 18 points, so the sequential stage takes up 20\% of the total budget.

Sobol', MCD, and MaxPro designs are used for comparison. The MCD is constructed from an orthogonal array $OA(72, 6^5(36), 2)$, where columns with six levels can be collapsed to two or three levels. Figure \ref{fig:CV_test_error} plots the cross-validation error and test error (evaluated on all rows not in the training set) of the 18 sequential points acquired for different initial designs. Although all designs achieve similar optimal performance, when using MCD and MaxPro, the active learning procedure still proposes many suboptimal points. In contrast, with MOFAT as the initial design, almost all hyperparameter settings in the sequential stage have a small test error. 

\begin{figure}[tb]
\centering
\includegraphics[width=\textwidth]{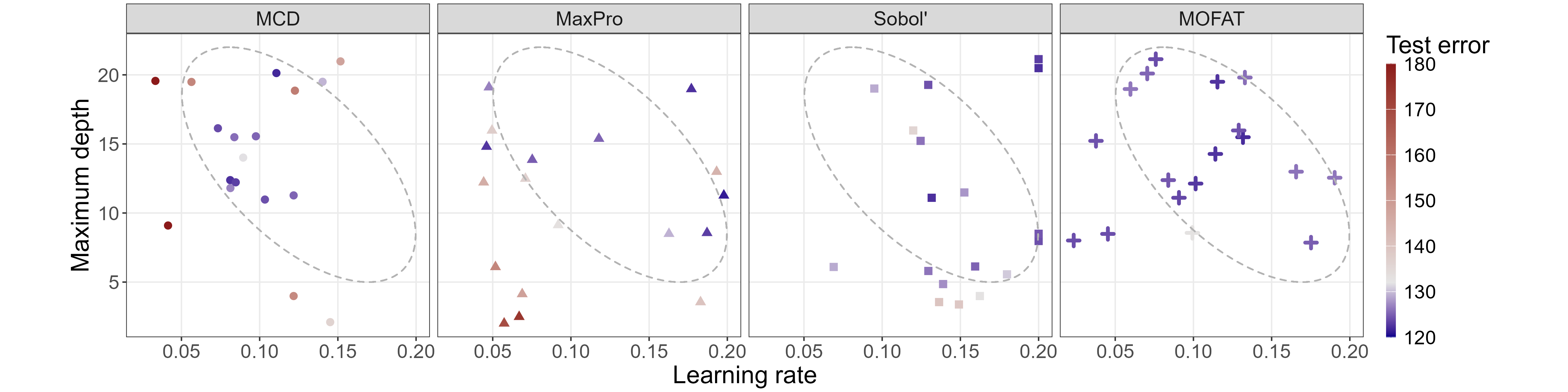}
\caption{Learning rates and maximum depths proposed during the sequential design. The ellipse marks the region with low error. Points are jittered for visibility.}
\label{fig:sequential_points}
\end{figure}

Figure \ref{fig:sequential_points} plots the maximum depth and learning rate (second and third most influential, after the type of boosting) of the sequential points. We can see that many different settings achieve a small error, indicating that the problem is multi-modal. Nevertheless, the desirable settings lie within a certain region. MOFAT can identify the driving forces of improving the objective, allowing most sequential points to lie within this region. See \citet{Song_Joseph} for further discussions on the benefits of using screening designs for active learning. On the other hand, with other designs as the initial design, we still invest in exploring settings with larger errors in the sequential stage, such as trees with shallow depth (the points near the bottom of the plots). In this example, MOFAT efficiently finds a ``basket'' of diverse solutions with good performance, which can be useful in downstream decision-making \citep{Joseph_Dasgupta_Tuo_Wu_2015, miller2025expecteddiverseutilityedu}. For example, the maximum tree depth can be 12, 15, or 20, but we can choose 12 to control the cost of training and prediction. 

\section{Conclusion} \label{sec:conclusion}

Many existing designs for computationally expensive black-box models accommodate both qualitative and quantitative factors. However, none of the designs emphasize identifying important factors. This article bridges the gap by extending the work of \citet{Xiao_Joseph_Ray_2023} on continuous factors to incorporate nominal, ordinal, and discrete-numeric factors into the MOFAT design. We propose three properties for constructing an optimal screening design, with different implications for different types of factors. We also provide efficient implementations available through the R package \texttt{MOFAT}. The design is flexible and economical in run size, being able to screen a large number of factors using limited runs. We believe it can be useful in a wide variety of engineering and data science applications.

In \citet{Xiao_Joseph_Ray_2023} and this article, the designs are based on ``standard'' OFAT, where all changes are made from a base run. This structure helped us derive results such as Proposition \ref{proposition:projection}. One future direction is construction using ``strict OFAT'' (changes are made from the previous run) and its properties. Furthermore, the current implementation requires the total run size to be a multiple of $p + 1$. It would be interesting to explore strategies to further increase the flexibility of the run size. Finally, we demonstrated good surrogate modeling performance using the proposed design, but it remains open to investigate whether a mixture of OFAT and space-filling designs further reduces out-of-sample error.

\vspace{.25in}

\begin{center}
{\Large\bf Supplementary Materials}
\end{center}

\noindent The proofs of Theorem 1 and Propositions 3-5 are provided in the supplementary.

\vspace{.25in}

\if0\blind{
\begin{center}
{\Large\bf Acknowledgments}
\end{center}

\noindent This research is supported by  U.S. National Science Foundation grant DMS-2310637. The work was carried out as part of the first author's Ph.D. thesis at Georgia Tech.

\vspace{.25in}
} \fi

% \FloatBarrier
% \input{Appendix}

%\vspace{.25in}

%\pagebreak

\bibliography{bib}
\pagebreak

\end{document}